\documentclass{JHEP3} 

\usepackage{epsfig,multicol,bbm}
\usepackage{cite}

\newcommand\fverb{\setbox\fverbbox=\hbox\bgroup\verb}
\newcommand\fverbdo{\egroup\medskip\noindent%
            \fbox{\unhbox\fverbbox}\ }
\newcommand\fverbit{\egroup\item[\fbox{\unhbox\fverbbox}]}
\newbox\fverbbox
\newcommand{\nb}{\nonumber}

\newcommand{\ppUW}{$pp \to \Upsilon(1S) +W+X~$}

\title{  Associated production of $\Upsilon(1S)W$ at LHC in next-to-leading order QCD }

\author{Li Gang$^{(a)}$, Ma Wen-Gan$^{(b)}$, Song Mao$^{(a)}$ Zhang Ren-You$^{(b)}$ and Guo Jian-You$^{(a)}$ \\
$^a$School of Physics and Material Science, Anhui University, Hefei, Anhui 230039, P.R.China \\
$^b$Department of Modern Physics, University of Science and
Technology of China, Hefei, Anhui 230026, P.R.China \\
E-mail: \email{lig2008@mail.ustc.edu.cn}, \email{mawg@ustc.edu.cn}, \email{songmao@mail.ustc.edu.cn},
\email{zhangry@ustc.edu.cn}, \email{jianyou@ahu.edu.cn}. }

\vskip 15mm \abstract{ We investigate the complete next-to-leading
order (NLO) QCD corrections to $\Upsilon(1S) +W$ production at the LHC,
and predict theoretically the distribution of the $\Upsilon(1S)$
transverse momentum. We analyse the contributions from different
components up to the QCD NLO in the \ppUW process.
Our results show that the \ppUW process has a large production rate
and could be potentially detected at the LHC. We see that the
differential cross section for the $\Upsilon(1S)$ direct-production
at the LO is significantly enhanced by the QCD corrections, and the
$b\bar{b} [ ^3S_1^{(8)} ]$ contribution component dominates in the
whole plotted $p_T^{\Upsilon(1S)}$ region. We have also calculated the
$\Upsilon(1S)$ meson indirect-productions via feed-down decays of
$\Upsilon(2S)$, $\Upsilon(3S)$, $\chi_{b1}(1P)$, $\chi_{b2}(1P)$,
$\chi_{b1}(2P)$, and $\chi_{b2}(2P)$ mesons. We find that the
$\Upsilon(1S)$ indirect-productions can give important contributions
to the distribution of $p_T^{\Upsilon(1S)}$ for the \ppUW process
at the NLO. We conclude that the studying the $\Upsilon(1S)+W$
production at the LHC could provide an interesting opportunity in
testing the nonrelativistic QCD factorization formalism. }


\vfill \eject

\begin{document}

\par
\section{Introduction}
\par
The study of heavy quarkonium is one of the interesting
subjects in both theoretical and experimental physics, which offers
a good testing ground for investigating the Quantum Chromodynamics (QCD)
in both the perturbative and non-perturbative regimes. The factorization
formalism of nonrelativistic QCD (NRQCD) \cite{bbl} provides a
rigorous theoretical framework to describe the heavy-quarkonium
production and decay by separating the amplitude into two parts:
a short-distance part which can be expanded as a sum of terms in the
power of $\alpha_s$ and calculated perturbatively, and
long-distance matrix elements (LDMEs), which can be extracted from
experiment. The relative importance of the LDMEs can be estimated by
means of velocity scaling rules \cite{vrule}. A crucial feature of
the NRQCD is that the complete structure of the quarkonium Fock
space has been explicitly considered.

\par
By introducing the color-octet mechanism (COM), NRQCD has
successfully absorbed the infrared divergences into P-wave
\cite{bbl,p1a,p1b,p1c,p2} and D-wave \cite{d1,d2} decay widths of
heavy quarkonium, which can not be handled in the color-singlet
mechanism (CSM). The COM can successfully reconcile the orders of
magnitude discrepancies between the experimental data of $J/\psi$
production at the Tevatron \cite{comtev} and the CSM theoretical
predictions, even if it has been calculated up to the next-to-leading-order
(NLO). Substantial progress has been achieved in the calculation of high order QCD corrections to
charmonium production in order to clarify the validity and limitation of the NRQCD formulism.
Recently, the complete NLO calculation for polarization and unpolarization of direct
hadroproduction in NRQCD are presented by two groups \cite{pnloz,chao2011,Butenschoen:1105,pnlok}.
For unpolarization of direct $J/\psi$ production in Ref. \cite{pnloz},  two new linear
combinations of color-octet matrix elements are obtained from
the CDF data by the authors, and used to predict $J/\psi$ production at the LHC, which agree with the CMS data.
$J/\psi$ polarization puzzle also may be understood as the transverse components canceling between $^3S_1^{(8)}$ and $^3P_J^{(8)}$
 channels \cite{chao2011}. The other group performed a multi-process fit of the color-octet LDMEs for unpolarization of
 direct hadroproduction \cite{Butenschoen:1105}, and gave more global fitting results which
are consistent with NRQCD scaling rules. While, they find that the CDF polarization data (Run-II) \cite{cdf132001}
can not be interpreted by using these fit values \cite{pnlok}. Therefore, the existence of the COM is still under doubt
and far from being proven.

\par
The heavy quarkonium production associated with a gauge boson may be
a good process in testing the COM. In high energy collider
experiments, $W$, $Z$ and heavy quarkonium can be selected with purely
leptonic decay modes \cite{braaten}, which are particularly useful
in hadron collides, because they provide an enormous suppression of
the background. The $W$ or $Z$ production associated with a
$J/\psi$ has been studied \cite{B.A.Kniehl-lo}. In
reference \cite{B.A.Kniehl-lo} the authors gave a theoretical
prediction for associated production of heavy quarkonim and
electroweak bosons at hadron collider at the LO, and the numerical
results show that the transverse momentum distribution of heavy
quarkonium production is smaller in the CSM than that in the COM at the LO.
Recently, the NLO QCD corrections to $J/\psi + W$ and $J/\psi + Z$
associated productions at the LHC were calculated in nonrelativistic
QCD \cite{liw,songw}. The numerical results show
that the differential cross section at the LO is significantly
enhanced by the NLO QCD corrections.

\par
Since the mass of bottomonium is about 3 times of that of
charmonium, the value of $v^2$ is smaller in bottomonium
$(\thickapprox 0.1)$ than that in charmonium $(\thickapprox 0.3)$,
the expansion in $v^2$ should converge faster in bottomonium.
Therefore, the study of bottomonium may be a more suitable choice to
test the NRQCD factorization formalism. The $\Upsilon(1S)$
production associated with a $W$- or a $Z$-boson at hadron
colliders has been studied in Ref.\cite{braaten}. It was found that
the cross sections for $\Upsilon(1S)+W$ and $\Upsilon(1S)+Z$
productions are roughly $0.45~pb$ and $0.15~pb$ at the Tevatron and
roughly $4~pb$ and $2~pb$ at the LHC respectively, and the dominant
production mechanism is via the P-wave bottomonium production, which
is a $b\bar{b}$ bounded state of color-octet, and the
bottomonium subsequentially decays to $\Upsilon(1S)$. In experimental
aspect, this process has been studied once at the Tevatron
\cite{cdfuw}, but due to the limited sensitivity of the CDF
experiment in Run I, they didn't observe $\Upsilon(1S)$ mesons
production in association with $W$- or $Z$-bosons at the Tevatron.
Since the LHC has larger integrated luminosity and higher energy, we
expect LHC experiments can achieve the sensitivity sufficient to
observe the $\Upsilon(1S)$ plus a vector boson production. As we
know, the NLO QCD corrections to quarkonium production are normally
important in precision measurement
\cite{Butenschoen:1105,chao2011,pnloz,pnlok}. It is
necessary to provide a complete NLO theoretical prediction for
$\Upsilon(1S)$ mesons in association with $W$- or $Z$-boson
production, and compare the results with experiment.

\par
In this paper we calculate the full NLO QCD corrections to the
$\Upsilon(1S)$ production in association with a $W$-boson at the
LHC. For this process, only the $^3S_1$ color-octet (the $b\bar{b}
\left[ ^3S_1^{(8)} \right]$ Fock state) provides contribution at the
leading-order (LO). Even including the NLO QCD contributions up to
$\alpha^3_s$ and $v^7$ of LDMEs, there are only color-octets
$b\bar{b} \left[ ^1S_0^{(8)} \right]$, $c\bar{b} \left[ ^3S_1^{(8)}
\right]$ and $b\bar{b} \left[ ^3P_J^{(8)} \right]$ $(J=0,1,2)$, but
no color-singlet contribute to the direct prompt $\Upsilon(1S)$
production in association with a $W$ gauge boson process. Therefore,
the $\Upsilon(1S)+W$ production at the LHC can provide an ideal
ground to study the COM. Meanwhile, The production of
$\Upsilon(1S)+W$ may also provide lampposts for looking for new
physics \cite{braaten}. In the context of supersymmetric models
(SUSY), a charged Higgs boson lighter than about $180~GeV$ have
a sizable decay branching ratio into $\Upsilon(1S)W$ pairs if the ratio of
vacuum expectation values $tan(\beta)$ is small. Therefore, the
$\Upsilon(1S)+W$ production may be a very important background
process for searching for charged Higgs boson in SUSY models.

\par
The paper is organized as follows. We present the details of the
calculation for the process \ppUW in Section 2. In Section 3, we
give the numerical results and discussions at the LHC. Finally,
a short summary is given.

\vskip 5mm
\section{Calculation descriptions}
\par
The strategy of the NLO QCD calculations in this work is similar to
that used in the evaluations for the $pp \to J/\psi+W/Z$ process
\cite{liw,songw}. For simplicity, we sketch it in this section.

\par
The cross section for the $pp\to H+W$ process is expressed as
\begin{eqnarray}
\sigma\left(pp \to H+W+X\right)&=& \int
dx_1dx_2\sum_{i,j,n}\frac{\langle{\cal O}^{H}[n]\rangle}{
N_{col}(n)N_{pol}(n)}\hat{\sigma}\left(i+j\to
b\bar{b}[n]+W+X\right)\nb \\
&&\times \left[G_{i/A}(x_1,\mu_f)G_{j/B}(x_2,\mu_f)+(A
\leftrightarrow B )\right],
\end{eqnarray}
where $N_{col}(n)$ and $N_{pol}(n)$ refer to the numbers of colors
and polarization $b\bar{b}(n)$ states produced \cite{p2},
respectively. $\hat{\sigma}\left(i+j\to b\bar{b}[n]+W+X\right)$ is
the cross section for the short distance production of a $b\bar{b}$
quark pair in the color, spin and angular momentum state $n$ at the
parton level. $\langle{\cal O}^{H}[n]\rangle$ is the long distance
matrix element, which describes the hadronization of the $b\bar{b}$
quark pair into the observable bottomonium state $H$. $G_{i,j/A,B}$
are the parton distribution functions (PDFs), A and B refer to
incoming protons at the LHC. $i,j$ represents gluon and all possible light
quarks and anti-quarks
$(i,j=u,d,s,c,\bar{u},\bar{d},\bar{s},\bar{c})$. We ignore the
processes which are suppressed by a tiny Kobayashi-Maskawa factor.

\par
The short distance cross section for the production of a $b\bar{b}$
pair in a Fock state $n$, $\hat{\sigma}[i+j\to b\bar{b}[n]+W+X]$, is
calculated from the amplitudes which are obtained by applying
certain projectors onto the usual QCD amplitudes for open $b\bar{b}$
production. In the notations of Ref.\cite{p2} they are written as
\begin{eqnarray}
{\cal A}_{b\bar{b} [{}^1S_0^{(1/8)} ]} = {\rm Tr} \Big[ {\cal
C}_{1/8} \Pi_0 {\cal A} \Big]_{q=0}, \nonumber
\end{eqnarray}
\begin{eqnarray}
{\cal A}_{b\bar{b} [ ^3S_1^{(1/8)} ]} = {\cal E}_{\alpha} {\rm Tr}
\Big[ {\cal C}_{1/8} \Pi_{1}^{\alpha} {\cal A} \Big]_{q=0},
\nonumber
\end{eqnarray}
\begin{eqnarray}
{\cal A}_{b\bar{b} [ ^1P_1^{(1/8)} ]} = {\cal E}_{\alpha }
\frac{d}{dq_{\alpha}} {\rm Tr} \Big[ {\cal C}_{1/8} \Pi_0 {\cal A}
\Big]_{q=0}, \nonumber
\end{eqnarray}
\begin{eqnarray}
{\cal A}_{b\bar{b} [ ^3P_J^{(1/8)} ]} = {\cal E}_{\alpha
\beta}^{(J)} \frac{d}{dq_{\beta}} {\rm Tr} \Big[ {\cal C}_{(1/8)}
\Pi_1^{\alpha} {\cal A} \Big]_{q=0},
\end{eqnarray}
where ${\cal A}$ denotes the QCD amplitude with amputated bottom
spinors, the lower index $q$ represents the momentum of the bottom
quark in the $b\bar{b}$ rest frame. $\Pi_{0/1}$ are spin projectors
onto the spin singlet and spin triplet states. ${\cal C}_{1/8}$ are
color projectors onto the color singlet and color octet states.
${\cal E}_{\alpha}$ and ${\cal E}_{\alpha \beta}$ represent the
polarization vector and tensor of the $Q\bar{Q}$ states,
respectively.
\begin{figure}[!htb]
\begin{center}
\begin{tabular}{cc}
{\includegraphics[width=10cm]{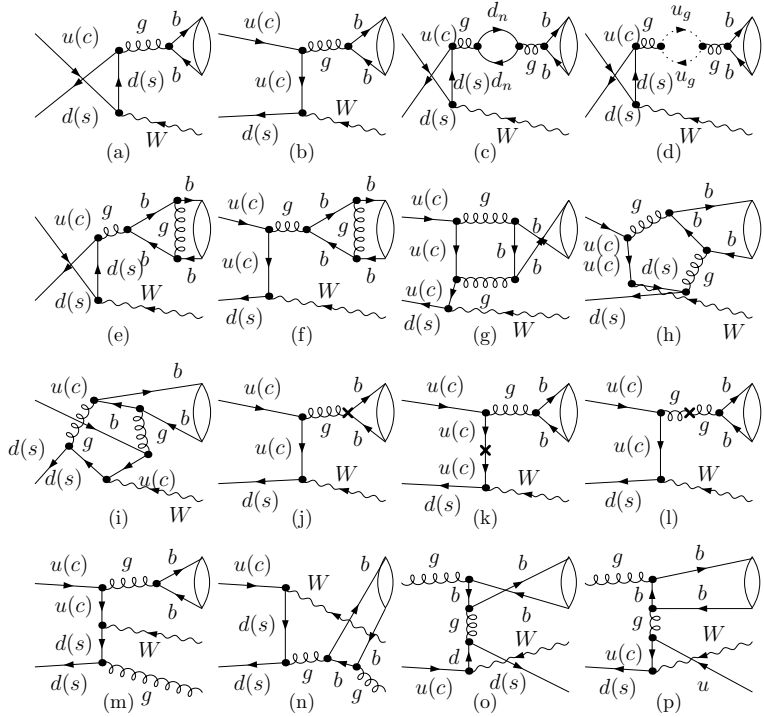}}
\end{tabular}
\end{center}
 \vspace*{-0.7cm}\caption{Some
representative Feynman diagrams for the process \ppUW at the LO and
up to QCD NLO.} \label{f1}
\end{figure}

\par
In the calculation for the LO of $\alpha_s$, only the Fock state
$^3S_1^{(8)}$ is involved. However, in considering the NLO QCD
corrections to $pp \to \Upsilon(1S)+W+X$ process, we should not only
consider the contribution from virtual corrections, but also the
contribution from real gluon/light-quark emission processes. We draw
some representative Feynman diagrams for the process \ppUW at the LO
and up to QCD NLO in Fig.1. The virtual corrections only occur in
connection with the $b\bar{b}$ Fock state $^3S_1^{(8)}$, but the
real gluon/light-quark emission process should involve
$^1S_0^{(8)}$, $^3S_1^{(8)}$ and $^3P_J^{(8)}$ Fock state
contributions. In our calculations, we adopt the dimensional
regularization (DR) method in $D = 4 - 2\epsilon$ dimensions to
isolate the ultraviolet (UV), soft infrared (IR), and collinear IR
singularities. There exist UV, soft, collinear and coulomb
singularities in virtual corrections. In order to remove the UV
divergences, we employ the modified minimal subtraction
($\overline{\rm MS}$) scheme to renormalize and eliminate UV
divergency. After applying the renormalization procedure the UV
divergences in the virtual correction part are canceled. The other
appeared singularities can also be analytically canceled when we
added all the NLO QCD contribution components together. In Fig.2, we
present the IR and Coulomb singularity structures and divergence
cancelation routes in the NLO QCD calculation for the $pp \to
\Upsilon(1S)+W+X$ process.
\begin{figure}[!htb]
 \vspace*{-0.5cm}
\begin{center}
\begin{tabular}{cc}
{\includegraphics[scale=1]{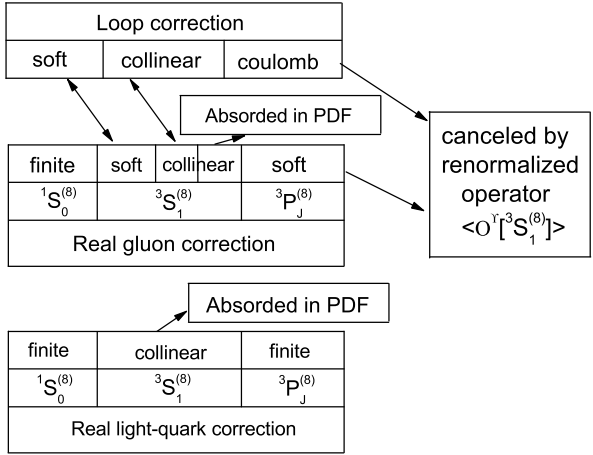}}
\end{tabular}
\end{center}
 \vspace*{-2.5cm}\caption{The IR and Coulomb
singularity structures in the NLO QCD calculations for the \ppUW
process. } \label{f2}
\end{figure}

\par
The real gluon process with $^1S_0^{(8)}$ Fock state is free of
divergence, while the contribution from the real gluon process with
$^3S_1^{(8)}$ Fock state has soft and collinear singularities, and
the real gluon processes with $^3P_J^{(8)}$ Fock states has soft
singularities. The soft divergences from the one-loop diagrams will
be canceled by similar singularities from the $^3S_1^{(8)}$ state
contribution of soft real gluon emission. As for the collinear
divergences from the $^3S_1^{(8)}$ state contribution of soft real
gluon emission corrections, part of it can be eliminated by
collinear singularities in virtual corrections, and the remaining
collinear divergences in real gluon corrections can be absorbed into
the PDFs. Nevertheless, it still contains coulomb singularities in
virtual corrections and soft singularities arising from the
$^3P_J^{(8)}$ state contribution of real gluon process, these
singularities is neither collinear nor infrared in the usual sense,
and it can only be eliminated in the spirit of the factorization
approach by taking the corresponding corrections to the operator
$<{\cal O}^{\Upsilon(1S)}[^3S_1^{(8)}]>$ into account. As for the
light-quark emission process, there only involve initial state
collinear divergences in $^3S_1^{(8)}$ state subprocess, and it also
can be absorbed into the PDFs. Finally, after adding all
contributions together, a result which is UV, soft, collinear,
Coulomb finite is obtained.

\par
We adopt the expressions in Ref.\cite{IRDV} to deal with the IR
divergences in Feynman integral functions, and apply the expressions
in Refs.\cite{OneTwoThree,Four,Five} to implement the numerical
evaluations for the IR safe parts of N-point integrals. In the
virtual correction calculation, we find that only Fig.1(e) and
Fig.1(f) induce coulomb singularities, and we use a small relative
velocity $v$ between $b$ and $\bar{b}$ to regularize them
\cite{coulomb}. The two cutoff phase space slicing method (TCPSS)
\cite{TCPSS} has been employed for dealing with the soft and
collinear singularities in real gluon/light-quark emission
corrections.

\vskip 10mm
\section{Numerical results and discussion}
\par
In the numerical calculations for the $pp\to \Upsilon(1S)+W+X$
process at the LHC, we take CTEQ6L1 PDFs with an one-loop running
$\alpha_s$ in the LO calculation and CTEQ6M PDFs with a two-loop
$\alpha_s$ in the NLO calculation\cite{CTEQ6}, and the corresponding
fitted value $\alpha_s(M_Z^2) = 0.130$ and $\alpha_s(M_Z^2) = 0.118$
are used for LO and NLO calculations respectively. We take the
number of active flavor as $n_f=4$. The relevant quark masses and
fine structure constant are taken as: $m_q=0,(q=u,d,c,s)$, $m_W=80.398~GeV$,
$m_b=m_{\Upsilon(1S)}/2=4.75~GeV$, $\alpha =1/137.036$. The
renormalization, factorization, and NRQCD scales are chosen as
$\mu_r=\mu_f=m_T$ and $\mu_{\Lambda}=m_b$, respectively, where
$m_T(\equiv\sqrt{(p_T^{\Upsilon(1S)})^2+4m_b^2})$ is the
$\Upsilon(1S)$ transverse mass. The NRQCD matrix elements of
$\Upsilon(1S)$ and other bottomoniums used in our calculations are list in Table \ref{tab1},
which determined from the CDF data \cite{km,u8m}.  In the calculation of real corrections,
we adopt the two-cutoff phase space slicing method \cite{TCPSS}.
The two phase space cutoffs $\delta_s$ and
$\delta_c$ are chosen as $\delta_s=10^{-3}$ and
$\delta_c=\delta_s/50$ as default choice. In checking the
independence of the final results on two cutoffs $\delta_s$ and
$\delta_c$, we find the invariance with the $\delta_s$ running from
$10^{-3}$ down to $10^{-4}$ within the error control (less than one
percent). Considering the validity of the NRQCD and perturbation
method, we restrict our results to the domain $p_T^{\Upsilon(1S)} >
3~{\rm GeV}$ and $|y_{\Upsilon(1S)}| < 3$.
\begin{table}
\begin{center}
\begin{tabular}{|c|ccc|}

  \hline
  H &$<\mathcal{O}^H_1>$ &$<\mathcal{O}^H[^3S_1^{(8)}]>$ & $<\mathcal{O}^H[^1S_0^{(8)}]>$  \\
  \hline
  \hline
  $\Upsilon(1S)$ &  & $15 \times 10^{-2}~GeV^3$ & $2.0 \times 10^{-2}~GeV^3$  \\
  $\Upsilon(2S)$ &  & $4.5 \times 10^{-2}~GeV^3$ & $0.6 \times 10^{-2}~GeV^3$  \\
  $\chi_0(1P)$   & $2.03~GeV^3$ & $4 \times 10^{-2}~GeV^3$ &   \\
  $\chi_0(2P)$   & $2.57~GeV^3$ & $6.5 \times 10^{-2}~GeV^3$ &   \\
  \hline

\end{tabular}
\end{center}
\begin{minipage}{15cm}
\caption{\label{tab1}  NRQCD matrix elements for bottomonium production, where colour-octet
matrix element $<\mathcal{O}^H[^1S_0^{(8)}]>=<\mathcal{O}^H[^3P_0^{(8)}]>/m^2_b$ has been
assumed for simplicity, and $m_b=4.75~GeV$. }

\end{minipage}
\end{table}

\begin{figure}[htbp]
\vspace*{-0.3cm} \centering
\includegraphics[scale=1]{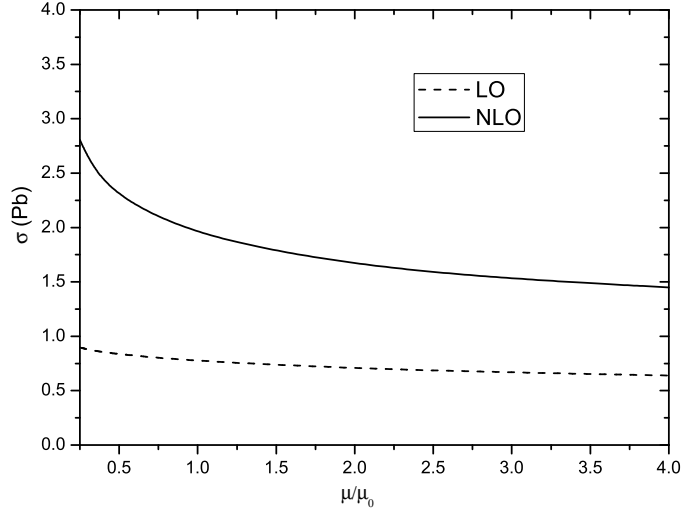}
\vspace*{-0.3cm} \centering \caption{\label{mu} The dependence of
the LO and the NLO QCD corrected cross sections for the process
\ppUW on the factorization scale and renormalization scale
($\mu$/$\mu_0$) at the $\sqrt{S}=14~TeV$ LHC where we define $\mu=\mu_f=\mu_r$ and
$\mu_0 = m_T$. }
\end{figure}

\par
The dependence of the cross section on the renormalization scale
$\mu_r$ and factorization scale $\mu_f$ induces uncertainty for
theoretical prediction. In Fig.\ref{mu}, the $\mu$ dependence of
the LO and the NLO QCD corrected cross sections with the constraints
of $p_t^{\Upsilon(1S)}
>3~GeV$ and $|y_{\Upsilon(1S)}| < 3$ for \ppUW process, is shown with our default
choice $\mu$ = $\mu_r$ = $\mu_f$ and the definition of $\mu_0=m_T$.
There $\mu$ varies from $\mu_0/4$ to $4 \mu_0$. Not like the usual
expectation, Fig.\ref{mu} shows that the NLO QCD correction can
not improve the LO scale independence for the \ppUW process at the
LHC. The related theoretical uncertainty varies from $+8.1\%$ to  $-8.6\%$ at the LO and
from $+18.5\%$ to $-13.8\%$ at the NLO when $\mu$ goes from $\mu_0/2$ to $2
\mu_0$. Actually, this behavior is similar to that of the $Wb\bar{b}$ production at the LHC (see Ref.\cite{wjj}).

\begin{figure}[!htb]
\vspace*{-1cm}
\begin{center}
\begin{tabular}{cc}
{\includegraphics[scale=0.65]{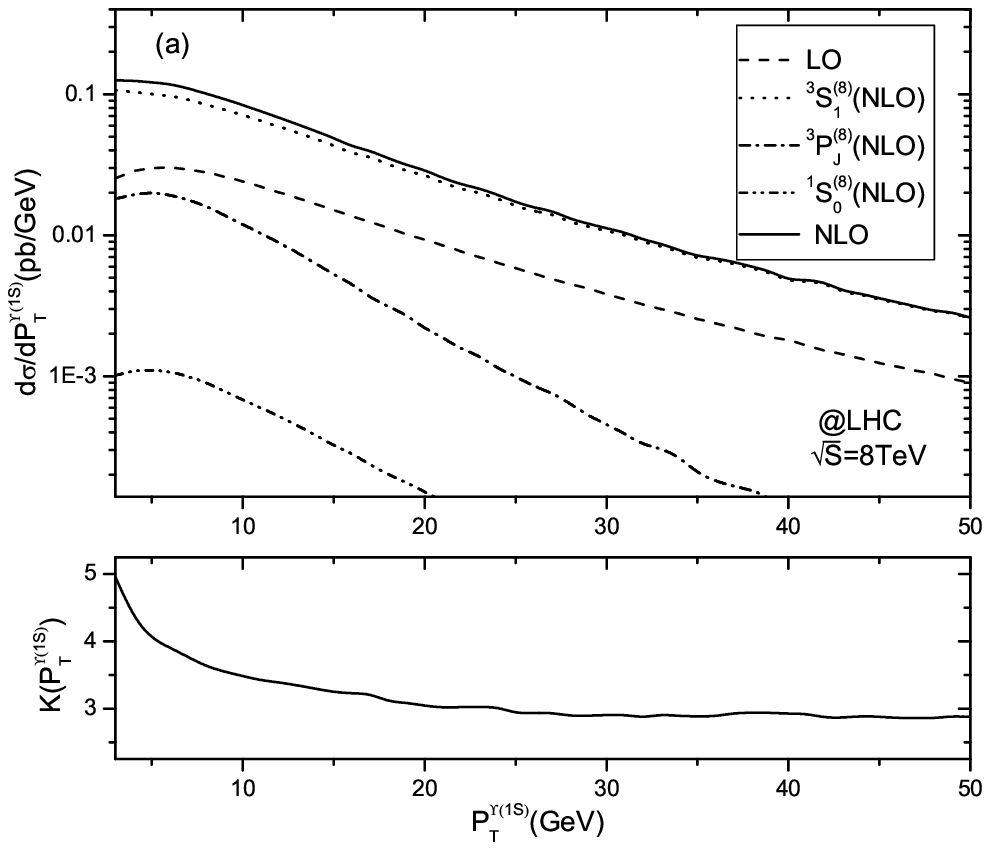}}
{\includegraphics[scale=0.65]{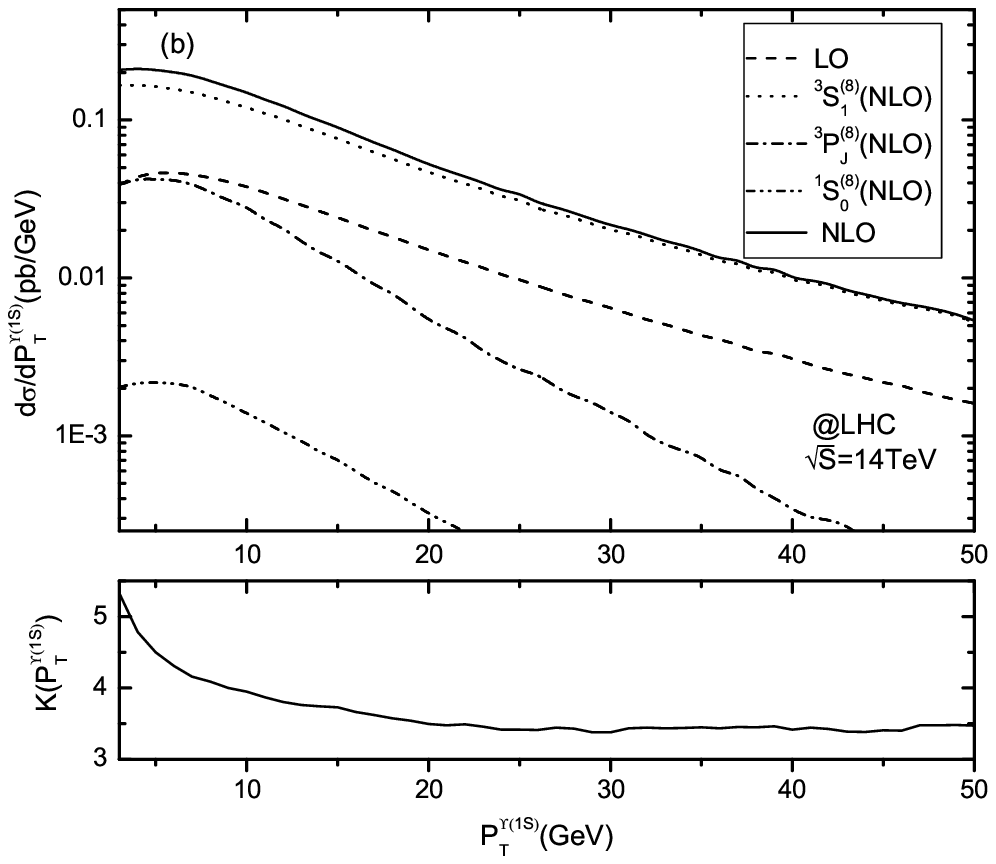}}
\end{tabular}
\end{center}
\vspace*{-1cm}
 \caption{The LO and NLO
QCD corrected distributions of $p_T^{\Upsilon(1S)}$ and the
corresponding K-factor for the direct $\Upsilon(1S) + W$ production.
(a) at the $\sqrt{S}=8~TeV$ LHC. (b) at the $\sqrt{S}=14~TeV$ LHC.} \label{upt}
\end{figure}

\par
In our numerical calculations, we add all the contributions of $pp\to
\Upsilon(1S)+W^++X$ and $pp\to \Upsilon(1S)+W^-+X$ together.
In Figs.\ref{upt}(a) and (b), we present the LO and the NLO QCD
corrected distributions of $p_T^{\Upsilon(1S)}$ and the corresponding
K-factors for the direct $\Upsilon(1S) + W$ production
at the $\sqrt{S}=8~TeV$ and the $\sqrt{S}=14~TeV$ LHC,
respectively. For comparison, we also depict the contributions from
the $b\bar{b} [ ^1S_0^{(8)} ]$, $b\bar{b} [ ^3S_1^{(8)} ]$ and
$b\bar{b} [ ^3P_J^{(8)}]$ Fock states in these figures. From the
figures we can see that the LO and the NLO corrected differential
cross sections are sensitive to $p_T$
of $\Upsilon(1S)$, and the LO differential cross section is
significantly enhanced by the NLO QCD corrections, and the $b\bar{b} [
^3S_1^{(8)} ]$ contribution dominates in the whole plotted
$p_T^{\Upsilon(1S)}$ region. In Fig.\ref{upt}(a), the NLO QCD
corrected differential cross section
of $p_T^{\Upsilon(1S)}$ at the $\sqrt{S}=8~TeV$
LHC decreases from $0.126~pb/GeV$ to $0.0026~pb/GeV$ as
$p_T^{\Upsilon(1S)}$ increases from $3$ to $50~GeV$. In the
range of $3~GeV < p_T^{\Upsilon(1S)} < 50~GeV$, the
$K$-factor, defined as $K =
\frac{d\sigma^{NLO}}{dp_T^{\Upsilon(1S)}}/\frac{d\sigma^{LO}}{dp_T^{\Upsilon(1S)}}$,
varies in the range of $[2.85,~4.96]$, and reaches its maximum when
$p_T^{\Upsilon(1S)} = 3~GeV$. From Fig.\ref{upt}(b), we find that
the NLO corrected differential cross section at the $\sqrt{S}=14 TeV$ LHC
in the whole plotted $p_T^{\Upsilon(1S)}$
region can be quantitatively beyond $3$ times of the corresponding LO
differential cross section, and reaches
its maximum $K =5.33$ when $p_T^{\Upsilon(1S)} = 3~GeV$. We can read out from
the figure that the NLO QCD corrected differential cross section varies in the range of
$[0.211,~0.0053]~pb/GeV$ when $p_T^{\Upsilon(1S)}$ runs from $3~GeV$ to $50~GeV$,
it reaches the maximum at the location of $p_T^{\Upsilon(1S)} = 4~GeV$.

\par
We have investigated the direct production of
$\Upsilon(1S)$ meson up to QCD NLO. The $\Upsilon(1S)$ meson can also be
produced indirectly via radiative or hadronic decays of heavier
bottomonium, such as $\Upsilon(2S)$, $\Upsilon(3S)$,
$\chi_{b1}(1P)$, $\chi_{b2}(1P)$, $\chi_{b1}(2P)$, and
$\chi_{b2}(2P)$ mesons. We know that these contributions are very
important in studying $\Upsilon(1S)$ meson production\cite{braaten}.
To calculate these feed-down contributions for $\Upsilon(1S)$ meson
production, the main problem is to determine the relation between
momentum of higher excited states and momentum of $\Upsilon(1S)$.
Following the method used in Ref.\cite{wk}, we ignored the momentum
shift when these excited states decay into $\Upsilon(1S)$. The cross
sections of these six feed-down production channels can be obtained
approximately by multiplying the direct-production cross section of
the intermediate bottomonium production with its decay branching
fraction to $\Upsilon(1S)$ meson. In our calculation, we ignored
the contributions aroused from $\chi_{b0}(1P,2P)$ feed-down into
$\Upsilon(1S)$ for the smallness of the transition branching ratios.

\par
The direct-production cross sections of the $\Upsilon(2S)$ and
$\Upsilon(3S)$ can be calculated by using analogous method to the
$\Upsilon(1S)$ production, and the relevant color-octet matrix elements of
$\Upsilon(2S)$ and $\Upsilon(3S)$ used in our calculations are
determined from the CDF data in Table \ref{tab1} \cite{km,u8m}. In the
calculation for the LO and the virtual corrections for the
direct-production cross sections of the $\chi_{b1}(1P)$,
$\chi_{b2}(1P)$, $\chi_{b1}(2P)$ and $\chi_{b2}(2P)$ mesons,
only the Fock state $^3S_1^{(8)}$ is involved. However, in considering the real
gluon/light-quark emission corrections process, we should not only
consider the contribution from $^3S_1^{(8)}$ Fock state
contribution, but also the contribution from $^3P_J^{(1)}$ Fock
state contributions. Based on NRQCD factorization
formalism\cite{p2,kk1}, we can handle these calculations by adopting
analogous method to the direct $\Upsilon(1S)$ production. For the NRQCD matrix
elements of $\chi_{bJ}$, the color-singlet matrix elements are taken from the
potential model calculation of Refs.\cite{km,u1ma,u1mb}, and the color-octet matrix
elements are determined from the CDF data \cite{km,u8m}. The
branching ratios of heavy bottomonium decays to $\Upsilon(1S)$ mesons
used in this paper are taken form Ref.\cite{pdg}.
\begin{figure}[!htb]
\vspace*{-1cm}
\begin{center}
\begin{tabular}{cc}
{\includegraphics[scale=0.7]{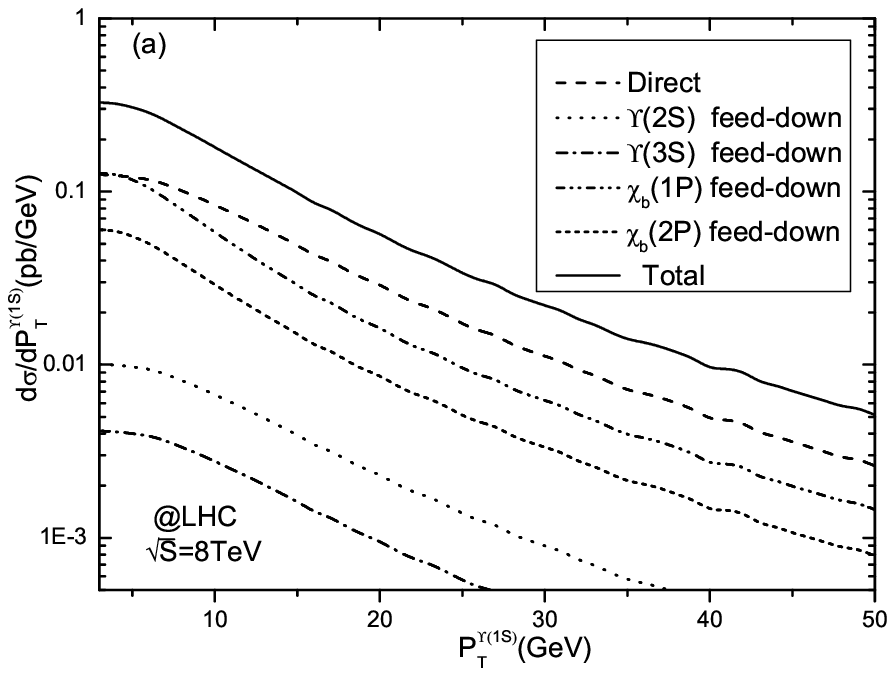}}
{\includegraphics[scale=0.7]{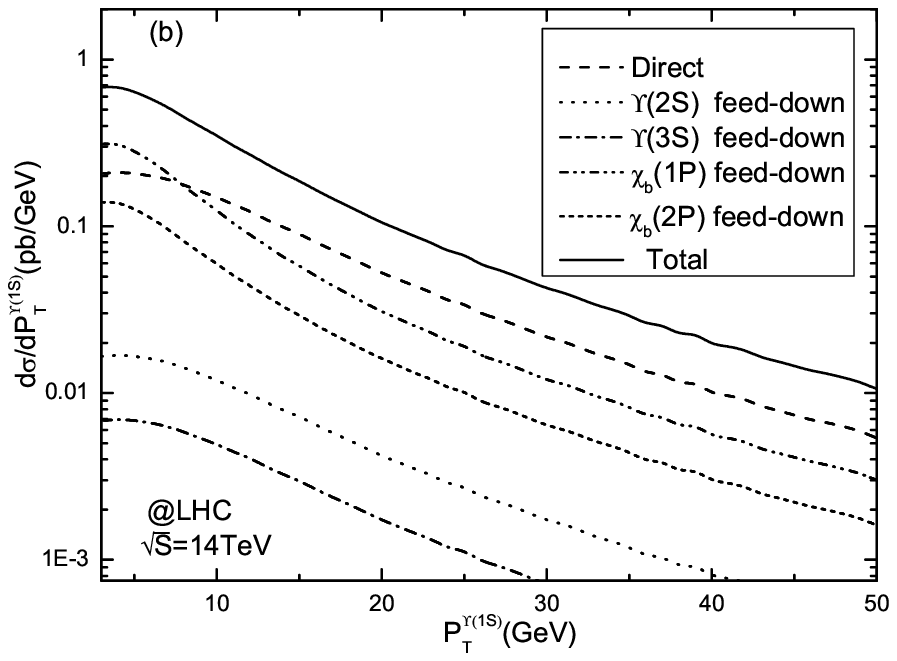}}
\end{tabular}
\end{center}
\vspace*{-1cm}
 \caption{The complete NLO
QCD corrected distributions of $p_T^{\Upsilon(1S)}$ for the \ppUW process.
(a) at the $\sqrt{S}=8~TeV$ LHC. (b) at the $\sqrt{S}=14~TeV$ LHC.}
\label{apt}
\end{figure}

\par
The complete NLO QCD corrected distribution of $p_T^{\Upsilon(1S)}$
for the \ppUW process, the contributions of the $\Upsilon(1S)$
direct-production and the $\Upsilon(1S)$ indirect-production by
$\Upsilon(2S)$, $\Upsilon(3S)$, $\chi_{b}(1P)$, and $\chi_{b}(2P)$
mesons feed-down decays at the LHC are illustrated in Fig.\ref{apt}. We can
see that the $\Upsilon(1S)$ indirect-production can
give important contribution to the distribution of $p_T^{\Upsilon(1S)}$
for the \ppUW process at the NLO. We find that when $p_T$ is
smaller than about $7~GeV$, the contribution from $\chi_{b}(1P)$
feed-down decays is about the same order of magnitude with the
contributions of $\Upsilon(1S)$ direct-production when $\sqrt{S}=8~TeV$,
particularly at the $\sqrt{S}=14~TeV$ LHC the productions of $\chi_{b}(1P)$
feed-down decays provide the main contribution to
$p_T^{\Upsilon(1S)}$ distribution for the \ppUW process. With the increment of the
$p_T^{\Upsilon(1S)}$, the contributions of $\Upsilon(1S)$
indirect-production contributions decrease quickly, and the
$\Upsilon(1S)$ direct-production contribution dominates at the
large $p_T$ region.

\vskip 5mm
\section{Summary}
\par
In this paper we investigate the complete NLO QCD correction of
$\Upsilon(1S) +W$ associated production at the LHC. This process is
an ideal platform for studying color-octet mechanism. We adopt the
dimensional regularization to deal with the UV and IR singularities
in our calculation. The Coulomb and soft singularities in P state
are isolated and absorbed into the NRQCD NLO corrected operator
$<{\cal O}^{\Upsilon(1S)}[^3S_1^{(8)}]>$. After
adding all contribution components together, we get the results with UV, IR,
Coulomb safety. We find that the production rate of the \ppUW
process is quite large, and this process has the potential to be
detected at the LHC. For the $\Upsilon(1S)$ direct-production, the
differential cross sections at the LO are significantly enhanced by
the QCD corrections, and the $b\bar{b} [ ^3S_1^{(8)} ]$ contribution
dominates in the range of $3~GeV < p_T^{\Upsilon(1s)} < 50~ GeV$. In
this paper, we also calculate the $\Upsilon(1S)$ meson
indirect-production via radiative or hadronic decays of
$\Upsilon(2S)$, $\Upsilon(3S)$, $\chi_{b1}(1P)$, $\chi_{b2}(1P)$,
$\chi_{b1}(2P)$, and $\chi_{b2}(2P)$ mesons. We find that the
$\Upsilon(1S)$ indirect-production contributions can give important
contribution to the distribution of $p_T^{\Upsilon(1S)}$ for the
\ppUW process at the NLO. In the lower $p_T$ region, the
contribution from indirect-production can give the main contribution
to $p_T^{\Upsilon(1S)}$ in the \ppUW process. We conclude that if the
$\Upsilon(1S)+W$ production is really detected at the LHC, it will
be useful for testing the NRQCD factorization formalism.

\vskip 5mm
\par
\noindent{\large\bf Acknowledgments:} This work was supported in
part by the National Natural Science Foundation of
China (No.11205003, No.11075150, No.11005101, No.11275190, No.11175001), the Key
Research Foundation of Education Ministry of Anhui Province of China
(No. KJ2012A021), and financed by the 211 Project of Anhui University (No.02303319).

\end{document}